\documentstyle[proc]{rspublic}
\begin{document}
\title[Exact Quantum Solutions of Extraordinary N-body Problems]{Exact Quantum  Solutions of Extraordinary N-body Problems}
\author[D. Lynden-Bell \& R.M. Lynden-Bell]{D. Lynden-Bell$^{1,2}$ and R.M. Lynden-Bell$^{1}$} 
\affiliation{$^1$School of Mathematics \& Physics, The Queen's
University, Belfast  BT7 1NN\\ $^2$Institute of Astronomy, Cambridge  CB3 0HA
and Clare College}
\maketitle
\label{firstpage}
\begin{abstract}
The wave functions of Boson and Fermion gases are known even when the
particles have harmonic interactions.  Here we generalise these
results by solving exactly the N-body Schr\"odinger equation for
potentials $V$ that can be any function of the sum of the squares of
the distances of the particles from one another in 3 dimensions.  For
the harmonic case that function is linear in $r^2$.  Explicit N-body
solutions are given when $U(r) = -2M \hbar^{-2} V(r) = \zeta r^{-1} -
\zeta_2 r^{-2}$.  Here $M$ is the sum of the masses and $r^2 = {1
\over 2} M^{-2} \Sigma \Sigma m_I m_J ({\bf x} _I - {\bf x}_J)^2$.  For
general $U(r)$ the solution is given in terms of the one or two body
problem with potential $U(r)$ in 3 dimensions.  

The degeneracies of the levels are derived for distinguishable
particles, for Bosons of spin zero and for spin 1/2 Fermions.  The
latter involve significant combinatorial analysis which may have
application to the shell model of atomic nuclei.

For large $N$ the Fermionic ground state gives the binding energy of a
degenerate white dwarf star treated as a giant atom with an N-body
wave function.

The N-body forces involved in these extraordinary N-body problems are
not the usual sums of two body interactions, but nor are forces between
quarks or molecules.  

Bose-Einstein condensation of particles in 3 dimensions interacting
via these strange potentials can be treated by this method.

\end{abstract}

\section{Introduction}

Exact Boson or Fermion solutions of the quantum N-body problem in
which every particle interacts with every other in three dimensions
are very rare.  They are almost as rare in classical mechanics
although Newton solved one in Principia (1687) (see Cajori 1934 \&
Chandrasekhar 1995) and there are also some very special solutions
such as Laplace's in which the three unequal masses describe ellipses
about their centre of mass while at each time they make an {\em
equilateral} triangle.  However Newton's solution was for all initial
conditions when the force on body $I$ due to body $J$ was of the form
$F_{IJ} = k m_I m_J ({\bf x}_J - {\bf x}_I)$.  Newton reduced this
problem to that of $N$ harmonic oscillators relative to the centre of
mass.  The quantum solution is similar to the $N$ oscillator solution
for solid state physics.  The potential energy of Newton's system is
$$V = {\textstyle {1 \over 2}} \! \sum_{\ \ I \ <} \! \sum_{\! \! \!  J} k m_I m_J ({\bf x} _I -
{\bf x} _J)^2 = {\textstyle {1 \over 2}} kM \Sigma m_I ({\bf x}_I -
{\overline {\bf x}})^2 \equiv {\textstyle {1 \over 2}} kM^2r^2\ . $$
Here we show that this solution may be generalised to systems in which
the total potential energy $V$ is any function of $r$.  We have
already explored these systems and their generalisations in Classical
Mechanics (Lynden-Bell \& Lynden-Bell 1999).  Except for Newton's
harmonic case all these systems give many-body forces in which the
force between any two bodies is approximately linear for separations
much less than the mean but with a coefficient that depends on that
current mean.  Despite the strange global nature of these force laws
they may be the only non-trivial quantum many-body problems that have
been solved exactly in 3 dimensions.  Only in Newton's harmonic case
do the forces reduce to simple pair-wise interactions.  It could be
argued that such global forces are unnatural, however, in some
respects the resulting behaviour mimics that found in Nature.  Ruth Lynden-Bell
(1995, 1996) showed that these systems can be used to give a simple
model of a phase transition that can be calculated even when $N$ is
small.  Also as we now show, such forces can be used to mimic some aspects of gravitation.

For a homogeneous sphere that may pulsate in radius, $a$, the
gravitational potential energy is $V = -{3 \over 5} GM^2/a$ and the
gravitational force on unit mass not outside $a$ is $-{4 \over 3} \pi
G \rho {\bf x}$.  The mean square radius of the sphere is $r^2 =
3a^2/5$ so $V = - \left( {3 /5} \right)^{3/2} GM^2/r$.  Now
forget real gravity but adopt this form of $V (r)$ for the
potential of one of our extraordinary N-body problems.  The force per
unit mass on any particle at ${\bf x}$ is given by
$$-M^{-1} \partial V/\partial {\bf x} = - \left({3\over
5}\right)^{3/2} GM {\bf x}/r^3 = - GM {\bf x}/a^3 = - {4 \over 3} \pi
G \rho {\bf x}$$ where ${\bf x}$ stands for any one of the ${\bf x}_I
- {\overline {\bf x}}$.

Thus for homogeneous spheres this choice of $V (r)$ in our
extraordinary N-body problem exactly mimics the effect of true gravity
both for global radial pulsations of the system and for the forces on
masses within it.  However if the system departs from homogeneity this
mimicry is no longer exact.  For inhomogeneous spherical systems the
true gravitational potential energy $V (r)$ can always be
written $-k GM^2/r$ with a $k$ that depends on the radial profile.  By
taking that to be the $V (r)$ in our extraordinary N-body
problem its Virial theorem will perfectly mimic that of the
gravitational problem but apart from the homogeneous case the forces
on the individual particles of which the system is composed will not
be the same in the mimic.  Outside gravitational theory the concept of
an effective potential is widely used in physics and chemistry, e.g.,
in the shell model of the nucleus, in quark-quark interactions at low
energy, and in modelling inter-molecular forces.  Now we have shown
that motion in these special potentials can be exactly calculated,
they will no doubt be used as approximations in those applications, as
well as many others.  Since the N-body wave-function is known exactly,
so is the correlation energy, but that may not be a useful general
guide to correlation because in our systems the net force on each
particle is directed radially toward the centre-of-mass whatever the
configuration of the other particles may be.  Furthermore in many real
systems the interaction between any two particles is strongest when
the particles are closest together while it is weakest for the systems
discussed here.  In spite of this it is possible to make systems that
are strongly repulsive when all particles try to come close together
and ones that behave like gravitating systems in the sense that the
overall radius obeys the Virial Theorem for a self-gravitating system.
Even without any repulsion the exclusion principle provides support
for systems of Fermions so with a $V \propto -GM^2/r$ appropriate for
gravity, we find configurations of White Dwarf type.

The N-body problems discussed here arose by direct generalisation of
Newton's work and so skipped the developments of the intervening
centuries.  We may nevertheless see how they fit into those
developments.  Liouville (1855) showed that if a system of $D$ degrees
of freedom had $D$ integrals of the motion whose mutual Poisson
Brackets vanished, then the remaining $D$ integrals of the motion
could be found as quadratures.  He also discovered a large class of
such separable systems while St\"ackel (1890) proved his necessary and
sufficient conditions for separability.  Whittaker (1904) gives a good
description of those works while he, Eddington (1915) and Eisenhart
(1934) helped to determine and classify such systems; De Zeeuw (1985)
gives a good historical introduction in his thesis paper.  Lynden-Bell
(1962) and Hall (1985) developed different ideas for finding classes
of systems with integrals or configuration invariants.  Carter (1968)
extended such results to the motion of charged particles in magnetic
fields in General Relativity.  Marshall and Wojciechowski (1988)
determined those potentials in $D$ dimensions for which the motion of
a classical particle separates in suitable coordinates and the
hyper-spherical potential of the systems discussed here can be viewed
as a highly degenerate member of their general $D$-dimensional
ellipsoidally separable potentials.  Evans (1990, 1991) has explored
systems that are superintegrable, having more than $D$ integrals for
$D$ degrees of freedom.  They separate in several different coordinate
systems and the integrals are the separation constants.  This is the
case for our hyperspherical systems (see Appendix).

In all these works separability was achieved by changing the
coordinates only.  The idea that separability might be achievable only
via canonical transformations involving the momenta as well as the
coordinates was not exploited.  Thus Kovalevski's top (1888) provided
an unexpected new system in which the separation was not of the
standard type.  Linear soluble systems were known which were not of the
simple separable type and Routh's (1877) Adam's prize essay on the
Stability of Motion gives a very thorough discussion.  Simple examples
of both linear (Freeman 1966) and non-linear problems (Vandervoort
1979, Contopoulos \& Vandervoort 1992) that need momentum dependent
transformations arose in stellar dynamics but it was only recently,
e.g., in the work of Sklyanin (1995) that more general ways of looking
for such systems were found.

Meanwhile a whole body of work based around Lax (1968) operator pairs and the
Inverse Scattering method showed that there were many previously
unsuspected exact solutions in both classical and quantum mechanics.
This field of endeavour is too large to be reviewed here so the reader
is referred to the review volumes Solitons (Bullough \& Caudrey
eds., 1980), Dynamical Systems VI Integrable Systems (Arnold ed.,
1995) and Soliton Theory: a survey of results (Fordy ed., 1990).
The connections between soluble models of N-body problems and field
theory are discussed in the book edited by Bazhanov \& Burden (1995).
In that volume quantum and classical integrable lattice models in one
dimension are considered by Bullough \& Timonen (1995) while two
dimensional models in statistical mechanics are discussed by Baxter
(1995).  

Prominent among many exactly soluble N-body models in one dimension
are the Toda lattice (Toda 1967, 1980) also discussed by Henon (1974)
and the Calogero model and its generalisations.  [Calogero (1971),
Sutherland (1971), Bullough \& Caudrey (1980), Olshanetsky \&
Perelomov (1995).]  In two dimensions Baxter's book (1982) and article
(1995) contain much of interest and certain solvable models in 3
dimensions have been proposed by Baxter (1986) and Bazhanov \& Baxter
(1992, 1993).  Probably the most prominent soluble field theory in 2
dimensions is that of Davey-Stewartson (1974), see Anker \& Freeman
(1978), the quantum version of which is considered in Pang et al.
(1990).

Sklyanin (1995) holds out the hope that all these soluble systems may
eventually be seen as special cases of the method of Separation of
Variables and produces some supporting evidence.

Although the above models of interacting systems of many Fermions or
Bosons can be solved exactly in one or two dimensions, the calculation
that follows may be the only exactly solved non-trivial three
dimensional N-body system yet known.  Furthermore suitable choices of
the function $V(r)$ will allow a study of the way the form of
interaction (albeit one of our strange global type) affects
Bose-Einstein condensation.  It should also prove possible by these
methods to study the effect of rotation on the condensation.  However
this paper is solely concerned with the solutions of Schr\"odinger's
equation with the correct symmetry in the wave function, so the
statistical mechanics and Bose-Einstein condensation displayed by
these models is not further discussed here.  With the somewhat more
realistic $\delta$ function interaction it has been studied previously
for one dimensional chains and their continuum limits, see, e.g.,
Bogoliubov et al. (1994) and Bullough \& Timonen (1998).  It is of
course the case that all soluble models are exceptional and a good
example of the intricacies of non-soluble models was furnished by
Henon (1969).

\section{N-body Solutions of Schr\"odinger's Equation}

Let $m_I$ be the mass of the $I^{\rm th}$ particle and ${\bf x}_I$
its position vector.  Writing $M = \Sigma m_I$ for the total mass, the
centre of mass is given by
$${\overline {\bf x}} = \Sigma \mu_I {\bf x}_I \ , \eqno (2.1)$$
where $\mu_I = m_I/M$, which implies
$$\Sigma \mu_I = 1 \ . \eqno (2.2)$$

We define `mass weighted' coordinates relative to the centre of mass \break
${\bf r}_I = \mu^{1/2}_I \left( {\bf x}_I - {\overline {\bf x}}
\right)$ and an associated 3$N$ dimensional vector, ${\bf r}$, in the
space of all the ${\bf r}_I$ by
$${\bf r} = \left( {\bf r}_1 , {\bf r}_2, {\bf r}_3 \ldots 
{\bf r}_N \right) \ . $$
The length of ${\bf r}$ is the mass weighted r.m.s. radius of the
system since
$$r^2 = \Sigma \mu_I \left( {\bf x}_I - {\overline {\bf x}} \right)^2 \
. \eqno (2.3)$$
\vfill 
\eject

\noindent
This expression may be rewritten in terms of the mutual separations of
the particles since
$$\displaylines{r^2 = \sum_I \mu_I \left( {\bf x}_I - {\overline {\bf
x}} \right) \cdot {\bf x}_I = \sum_I \sum_J \mu _I \mu _J \left( {\bf
x}_I - {\bf x}_J \right) \cdot {\bf x}_I = \cr \hfill{} = \sum_J
\sum_I \mu_J \mu_I \left ( {\bf x}_J - {\bf x}_I\right) \cdot {\bf x}_J \
,\hfill{}\cr} $$ 
and by adding the last two expressions and halving
the result
$$r^2 = {\textstyle {1 \over 2}} \sum_I \sum_J \mu_I \mu_J ({\bf x}_I - {\bf
x}_J)^2 = \sum_{\ \ I \ <} \! \sum_{\! \! \! J} \mu_I \mu_J ({\bf x}_I
- {\bf x}_J)^2 \ . \eqno (2.4)$$
In practice the ${\bf r}$ vector is constrained by the fact that the
centre of mass is at the origin so
$$\Sigma \mu^{1/2}_I {\bf r}_I = \Sigma \mu_I ({\bf x}_I - {\overline
{\bf x}}) = 0 \ . \eqno (2.5)$$

We define three mutually orthogonal unit vectors ${\hat {\bf X}}$,
${\hat {\bf Y}}$ and ${\hat {\bf Z}}$ in our 3$N$ space by
$$
{\hat{\bf X}} = \left(\mu^{1/2}_1,\, 0,\, 0,\, \mu^{1/2}_2 ,\, 0 ,\, 0
\ldots \mu^{1/2}_N, \, 0 , \, 0\right)$$
$$
{\hat{\bf Y}}  =  \left(0,\,  \mu^{1/2}_1,\, 0,\, 0,\, \mu^{1/2}_2,\, 0 \ldots 0,\,
\mu^{1/2}_N,\, 0\right) \eqno  (2.6)$$
$$
\ \ \ {\hat {\bf Z}}  =  \left(0,\, 0,\, \mu^{1/2}_1,\, 0,\, 0,\, \mu^{1/2}_2 \ldots 0,\,
0,\, \mu^{1/2}_N \right)\ . $$
Then the constraints (2.5) can be rewritten
$${\hat {\bf X}} \cdot {\bf r} = {\hat {\bf Y}} \cdot {\bf r} = {\hat
{\bf Z}} \cdot {\bf r} = 0 \eqno (2.7)$$
which show that ${\bf r}$ is confined to three hyperplanes through the
origin.  Defining ${\hat {\bf r}} = {\bf r}/r$ then ${\hat {\bf r}}$
lies on the unit 3$N$ sphere $|{\hat {\bf r}}|^2 = 1$ but the ${\hat
{\bf X}}$ constraint confines it to the intersection of that sphere
with the hyperplane ${\hat {\bf X}}\cdot {\bf r} = 0$, which is a
sphere in 3$N$-1 space; similarly the ${\hat {\bf Y}}$ constraint
leaves it on the intersection of that 3$N$-1 sphere with the
hyperplane ${\hat{\bf Y}} \cdot {\bf r} = 0$, which is a 3$N$-2 sphere
and the third constraint leaves it on the 3$N$-3 sphere orthogonal to
${\hat {\bf X}}$, ${\hat{\bf Y}}$ and ${\hat {\bf Z}}$.

We are concerned with the N-body problems whose potential energies,
$V$, are functions of the magnitude $r$ only, so Schr\"odinger's
equation takes the form
$$-{\textstyle{1 \over 2}} \hbar^2 \sum_I m^{-1}_I {\partial^2 \psi
\over \partial {\bf x}_I \cdot \partial {\bf x}_I} + V \psi = E_T \psi
\ . \eqno (2.8)$$
The key to solving this problem lies in the right choice of
coordinates.  In what follows upper case indices run over particle
labels while lower case indices run over coordinate-vector components.

\subsection{Separation of ${\overline x}$}

Let $R_{ij}$ be an orthogonal unit $3N \times 3N$ rotation matrix
which rotates the basis vectors of our $3N$ space so that ${\hat{\bf
X}}$, ${\hat{\bf Y}}$ and ${\hat{\bf Z}}$ are the last three of the
new orthogonal basis vectors.  Thus with two alternative notations and
assuming the summation convention over lower case indices only,
$$q_i = R_{ij} r_j = \sum_J {\bf R} _{iJ} \cdot {\bf r}_J \ , \eqno
(2.9)$$
with
$$q_{3N-2} = X \ , \ q_{3N-1} = Y \ , \ q_{3N} = Z \ . \eqno (2.10)$$
Note
$$R_{ij}R_{kj} = \delta _{ik} \ {\rm and} \ \sum_J {\bf R}_{iJ} \cdot
{\bf R}_{kJ} = \delta_{ik} \ . \eqno (2.11)$$

Let $a$ run from 1 to 3$N$-3 (rather than from 1 to $3N$).  Then the
$q_a$ together with the coordinates ${\overline {\bf x}}$ form a
complete set of independent orthogonal coordinates for our system.

We shall need the partial derivatives from (2.1) and below (2.2)
$$\partial {\overline{\bf x}}/\partial {\bf x}_I = \mu_I
{\underline{\mbox{\boldmath$ \delta$}}} \ , \eqno (2.12)$$
$$\partial {\bf r}_J/\partial {\bf x}_I = \sqrt{\mu_J} \left
( \delta_{JI} - \mu_I \right) {\underline{\mbox{\boldmath$ \delta$}}} \ , 
\eqno (2.13)$$
where $
{\underline{\mbox{\boldmath$ \delta$}}}$ is the unit $3 \times 3$
matrix.  The centre of mass motion will separate so our wave functions
may be taken in the form $\psi = {\overline \psi} ({\overline {\bf
x}}) \widetilde \psi (q _a \ldots )$ so
$${\partial \psi \over \partial {\bf x}_I} = {\partial {\overline
\psi} \over \partial {\overline {\bf x}}}\, \mu_I \widetilde \psi +
{\overline \psi} \, {\partial \widetilde \psi \over \partial {\bf x}_I} \ .
\eqno (2.14)$$

For Schr\"odinger's equation we shall need
$$\sum_I \mu^{-1}_I\, {\partial ^2 \psi \over \partial {\bf x}_I \cdot
\partial {\bf x}_I} = {\partial^2 \overline \psi \over \partial
{\overline{\bf x}} \cdot \partial {\overline {\bf x}}} \, \widetilde
\psi + \overline \psi \Sigma \mu^{-1}_I {\partial^2 \widetilde \psi
\over \partial {\bf x}_I \cdot {\bf x}_I} \ , \eqno (2.15)$$ where the
cross derivative term has vanished because $\widetilde \psi$ only
involves differences of the coordinates $x_I$ so $\Sigma
\partial \widetilde \psi/\partial {\bf x}_I = 0$, i.e., $\widetilde
\psi$ is independent of where the system-as-a-whole is.

To evaluate the second term we need from (2.9) and (2.13)
$${\partial \widetilde \psi\over \partial {\bf x}_I} = \sum_K {\partial
\widetilde \psi \over \partial q_j} \, {\partial q_j \over \partial
{\bf r}_K} \cdot {\partial {\bf r}_K \over \partial {\bf x}_I} =
{\partial \widetilde \psi \over \partial q_j} \left( {\bf R}_{jI}
\sqrt{\mu_I} - \sum_K \sqrt{\mu_K} \mu_I {\bf R}_{jK} \right) \
. \eqno (2.16)$$
We check that indeed $\Sigma \partial \widetilde \psi / \partial {\bf x}_I =
0$ by summing this over $I$ and noting that the two sums cancel because $\Sigma  \mu_I = 1$.  We now proceed to the last term in
Schr\"odinger's equation (2.15)
$$\sum_I {1 \over \mu_I}\, {\partial^2 \widetilde \psi \over \partial
{\bf x}_I \cdot \partial {\bf x}_I} = {\partial^2 \widetilde \psi
\over \partial q_\ell \partial q_j }\, \sum_I {\bf R}_{jI} \cdot {\bf
R}_{\ell I} - \sum_K \sum_I {\partial^2 \widetilde \psi \over \partial q_j
\cdot \partial {\bf x}_I} \sqrt{\mu_K} {\bf R}_{jK} \ .$$ 
But the last term involves $${\partial \over \partial q_j} \left
( \sum_I {\partial \widetilde \psi \over \partial {\bf x}_I} \right)$$
which is zero and $$\sum_I {\bf R}_{jI} \cdot {\bf R}_{\ell I} =
\delta_{j\ell}$$ because $\bf R$ is an orthogonal matrix.  Hence we have the
desired expression
$$\sum_I \mu^{-1}_I {\partial^2 \psi \over \partial {\bf x}_I \cdot
\partial {\bf x}_I} = {\partial^2 {\overline \psi} \over \partial
{\overline {\bf x}} \cdot \partial {\overline {\bf x}}} {\widetilde
\psi} + {\overline \psi} {\partial^2 {\widetilde \psi} \over \partial
q_a \partial q_a} \ . \eqno (2.17)$$
$a$ has replaced $j$ in the final term because $\widetilde \psi$
is only dependent on the first 3$N$-3 of the $q_j$, and we remember
that $q_aq_a = r^2$ since $X$, $Y$ and $Z$ are all zero.

On division by $\psi$ Schr\"odinger's equation now takes the form
$$-{\textstyle {1 \over 2}} \hbar^2 M^{-1} \left( {1 \over {\overline
\psi}} \, {\partial^2 {\overline \psi} \over \partial {\overline {\bf
x}} \cdot \partial {\overline {\bf x}}} + {1 \over {\widetilde \psi}}
\, {\partial^2 {\widetilde \psi} \over \partial {\bf q} \cdot \partial
{\bf q}} \right) + V(r) = E_T \ , \eqno (2.18)$$
where ${\bf q}$ stands for the $3(N-1)$ vector $q_a$.  The equation
clearly separates with the final three terms dependent on the $q_a$
only and the first dependent on ${\overline {\bf x}}$ only, so it must
be constant.  Without loss of generality we can take the total
momentum to be $\hbar {\bf K}$.  Then ${\overline \psi} = \exp (i{\bf
K} \cdot {\overline {\bf x}})$ and writing $E = E_T - {1 \over 2}
\hbar^2 K^2/M$ we find 
$$- {\hbar^2 \over 2M} \, {\partial^2 {\widetilde \psi} \over \partial
{\bf q} \cdot \partial {\bf q}} + V(r) {\widetilde \psi} = E
{\widetilde \psi} \eqno (2.19)$$
where ${\bf q} \cdot {\bf q} = r^2$.  

\subsection{Separation of Angular Coordinates}

Equation (2.19) clearly separates again in hyperspherical polar
coordinates but it is simplest to write them symbolically by putting
${\bf q} = r {\hat {\bf r}}$ and regarding $r$ as independent of the
angular coordinates ${\hat {\bf r}}$.  We need the partial
differentials
$$\partial r/\partial {\bf q} = {\hat {\bf r}} \eqno (2.20)$$
$$\partial {\hat {\bf r}}/ \partial {\bf q} = \partial /\partial {\bf
q} \left( {\bf q}/r\right) =
r^{-1} ({\underline {\mbox {\boldmath $\delta$}}} - {\hat {\bf r}}
{\hat {\bf r}} ) \eqno (2.21)$$
and by writing ${\hat {\bf r}} = {\bf q}/r$ and using (2.20)
$$ \partial / \partial {\bf q} \cdot {\hat {\bf r}} = r^{-1} (3N-4) \
. \eqno (2.22)$$
We write ${\widetilde \psi} = \psi_r (r) {\hat {\psi}} ({\hat {\bf r}})$ and notice
that ${\hat \psi}$ is constant on radial lines so that ${\hat {\bf r}}
\cdot \partial {\hat \psi}/\partial {\hat {\bf r}} = 0$.  Then
$${\partial {\widetilde \psi} \over \partial {\bf q}} = {\partial \psi_r \over
\partial r} \, {\hat {\bf r}} {\hat \psi} + r^{-1} \psi_r {\partial {\hat \psi}
\over \partial {\hat {\bf r}}} \eqno (2.23)$$
and
$${\partial^2 {\widetilde \psi} \over \partial {\bf q} \cdot \partial {\bf q}} = {\partial
^2 \psi_r \over \partial r^2} \, {\hat \psi} + (3N-4) r^{-1} {\partial \psi_r
\over \partial r} \, {\hat \psi} + r^{-2} \psi_r {\partial^2 {\hat \psi} \over
\partial {\hat {\bf r}} \cdot \partial {\hat {\bf r}}} \eqno (2.24)$$
dividing by $\psi$ Schr\"odinger's equation now takes the form
$$\psi^{-1}_r \left[ r^2 \partial ^2 \psi_r / \partial r^2 + (3N-4)r
\partial \psi_r/\partial r \right] - \alpha^2 r^2 + U(r) r^2 = {\hat
{\psi}}^{-1} A ({\hat {\psi}}) \ . \eqno (2.25)$$ $\alpha$ is given by
$\alpha^2 = -2ME \hbar^{-2}$ and $U(r) = -2MV(r) \hbar^{-2}$.  The
angular operator $A$ (the hyper-angular-momentum operator) is given by
$$A ({\hat \psi}) = - \partial^2 {\hat \psi}/\partial
{\hat {\bf r}} \cdot \partial {\hat {\bf r}} \ . \eqno (2.26)$$

The angular operator $A$ also appears in the generalised $\nabla ^2$
in 3$N$-3 dimensions viz
${\partial^2 / \partial {\bf q} \cdot \partial {\bf q}}$,
so we shall first study the hyper-spherically symmetric solutions of
$\partial^2 \chi / \partial {\bf q} \cdot \partial {\bf q} = 0$.
Evidently
$${d^2 \chi \over dr^2} + {(3N-4) \over r} \, {d \chi \over d r} = 0$$
so $d \chi/dr = Cr^{-(3N-4)}$ therefore $\chi = B r^{-(3N-5)}$ where
we have omitted a constant of integration which is irrelevant to our
purpose.  We see this is correct by considering the first non-trivial
case in which terms other than ${\overline {\bf x}}$ are involved
which is $N=2$.  Then the space of the $q_a$ is 3 dimensional and the
elementary solution has $\chi = Br^{-1}$.  Now our generalised
$\nabla^2$ knows no particular origin so if $\chi$ is a solution for
$r \neq0$ then so is ${\widetilde \chi} = B |{\bf r} - {\bf r}_0 |
^{-(3N-5)}$ for ${\bf r} \neq {\bf r}_0$.  We expand such solutions
both for $|{\bf r}| < |{\bf r}_0|$ and for $|{\bf r}| > |{\bf r}_0|$ in powers
and the coefficients of these powers are $Y_L ({\hat {\bf r}})$
hyperspherical harmonics, just as they are in 3 dimensions.  In
particular the $L^{\rm th}$ harmonics have a power in $r$ of either
$r^L$ or $r^{-L-(3N-5)}$.  Looking for solutions of the form
$r^Lf({\hat {\bf r}})$ to our generalised $\nabla^2 =0$ we see that in $D=3N-3$ dimensions
$$r^2 {d^2 \over dr^2} \left(r^Lf\right) + (D-1) r {d \over dr}
\left( r^Lf \right) = r^L A(f) \ ,$$
i.e.,
$$L(L + D-2)\,  f = A(f) \eqno (2.27)$$
and hence the eigenvalues of $A(f)$ are $L(L+3N-5)$.
Notice that for a two body problem this reduces to the $L (L+1)$ in 3
dimensions that we know so well.  We shall return later to look for
the degeneracies of these different eigenstates but for solving
Schr\"odinger's equation the eigenvalues are sufficient.  For the
detailed separation of the $3N-4$ angular coordinates each in turn see
the Appendix.

\subsection{The Radial Equation}

Schr\"odinger's equation (2.25) now reads 
$$d^2 \psi_r/dr^2 + (3N-4)r^{-1} d\psi_r/dr - [\alpha^2 - U(r) +L
(L+3N-5)r^{-2}] \psi_r = 0 \ . \eqno (2.28)$$
Now in the corresponding classical N-body problem we showed that the
solution for the radial pulsations of the whole N-body system could
be found in terms of the radial pulsation of the corresponding two
body problem with the same $U(r)$.  With $N=2$ we have the usual
Schr\"odinger equation for a spherical potential with angular momentum
$\ell$
$$d^2 \psi_2/dr^2 + 2r^{-1}d\psi_2/dr - (\alpha^2 - U(r) + \ell (\ell
+1) r^{-2}) \psi_2 = 0 \ . $$
We shall suppose that this problem has been solved and the
corresponding eigenvalues and eigenfunctions are known.  Now put
$\psi_2 = r^{\beta} \chi_2$; then $\chi_2$ obeys
$$\chi_2 '' + (2 \beta +2) r^{-1} \chi'_2 - \left[ \alpha^2 - U(r) + \ell
(\ell + 1) - \beta (\beta +1) \right] \chi_2 = 0 \ . \eqno (2.29)$$
Writing $\ell (\ell +1) - \beta (\beta +1) = (\ell - \beta) (\ell + \beta
+1)$  one notices that putting
$$\beta = {\textstyle {3 \over 2}} (N-2) \eqno (2.30)$$ 
and $\ell = L+
\beta$,
transforms equation (2.29) into precisely (2.28).  We deduce the
surprising theorem below:

{\em Theorem}:  If the energy levels of the usual Schr\"odinger
equation in 3 dimensions are $E(P, \ell)$, where $P= 1,2,3 \ldots$ is
the radial quantum number and $\ell$ is the angular momentum quantum
number, then the energy levels of the N-body  problem with the `same'
`potential' $U(r)$ are $E \left(P, L+ {\textstyle {3 \over 2}}
(N-2)\right)$.  Furthermore the radial parts of their wave functions
are related by $\psi_r (P, L, r) = r^{-{3 \over 2} (N-2)} \psi_2 \left(P, L
+ {3 \over 2} (N-2), r \right)$.

Notice that the ground states are not the same because $L+{3 \over
2}(N-2)$ can not be zero for $N>2$.  Thus the ground state of the
N-body system corresponds to an $\ell$ of ${3 \over 2} (N-2)$ which
can be a high angular momentum state of the two body problem.
We see at once that the N-body problem will have proper energy levels
even if strongly {\em attractive} $r^{-2}$ potentials are added to $U$
just because of this effective increase in the $\ell$ of the ground
state.

The theorem above is very powerful in that it enables us to use
everything that is known about the solutions to the normal
Schr\"odinger equation in spherical potentials and transform it into
knowledge of our N-body problems.  As is well known, not only the
generalised Kepler potential $U(r)\! =\! \zeta r^{-1}\! -\! \zeta_2
r^{-2}$, but also similarly generalised square well potentials, and
the $U \propto \delta (r-r_0)$, $\delta$-function potentials as well
as oscillator potentials $U(r) = -{1 \over 2} \kappa^2 r^2 - \zeta_2
r^{-2}$ all have pretty solutions for the eigenvalues and
eigenfunctions.  All this takes over directly.  However, when
$\zeta_2$ is so negative that the two body problem has no ground state
there is perhaps room for doubt as to whether we can use that
potential's higher angular momentum states for our N-body problem.
To allay any such doubts and provide a pretty illustration of the
truth of our general theorem, we now solve the N-body equation for the
generalised Kepler potential.

\subsection{Generalised Kepler Problem}

We have already shown that the wave function takes the form
$$\psi = \exp (i {\bf k} \cdot {\overline {\bf x}}) Y_L ({\hat {\bf
r}}) \psi_r (r)$$
where $Y_L$ may be given a further 3$N$-5 indices corresponding to
various components of $\bf L$.  We must now determine $\psi_r$.

In solving equation (2.28) we follow the standard method of solution
beautifully laid out in the book by Pauling \& Wilson (1935).  Setting
$\alpha r = \tilde r$ and $\zeta/\alpha = \tilde \zeta$ and keeping
only the terms that dominate at large $\tilde r$ we find
$$d^2 \psi_r / d \tilde r^2 - \psi_r \sim 0\ .$$
So the asymptotic solutions behave as $\exp \pm \, \tilde r$.  Only the
minus sign is acceptable so we write $\psi_r = \eta (\tilde r) \exp (-
\tilde r)$ and derive the equation for $\eta$ valid for all $\tilde
r$,
$$\eta '' + (3N\! - 4) \tilde r^{-1} \eta ' - 2 \eta' + \biggl \{ \left
[ {\tilde \zeta} - (3N\! \! - \! 4) \right] {\tilde r}^{-1}\!  - \!
\left[ \zeta _2 +L (L+3N\! \! -\! 5) \right] {\tilde r}^{-2} \biggr\}
\eta = 0 \; . \eqno (2.31)$$
We look for power series solutions of the form 
$$\eta = \sum^\infty_{p=0} a_p \tilde r ^{s+p}$$with $a_0 \neq 0$ and
find the recurrence relation
$$\displaylines{ \biggl\{ (p+s) (p + s + 3N -5) - \bigl[ \zeta_2 +L
(L+3N-5) \bigr] \biggr\} a_p = \cr \hfill = \left[ 2(p+s) + 3(N-2) -
\tilde \zeta \right] a_{p-1} \ . \hfill {(2.32)}}$$ The indicial
equation has $a_{-1} = p = 0 $ and yields a quadratic equation for $s$
$$s^2 + (3N -5) s - \left[ \zeta_2 + L (L+3N-5) \right] = 0 \ . \eqno
(2.33)$$
In the pure Kepler case with $\zeta_2=0$ this yields $s = L$ or
$-(L+3N-5)$ of which only the positive $s=L$ solution obeys the
boundary condition at the origin.  For general $\zeta_2$ the solutions
are (cf (2.30))
$$s = -{\textstyle {1 \over 2}} (3N-5) \pm \sqrt{ \left[ L+{\textstyle
{1 \over 2}} (3N-5) \right]^2 + \zeta_2} = - {\textstyle {1 \over 2}}
- \beta \pm \sqrt{ \left( {\textstyle{1 \over 2}} + L + \beta
\right)^2 + \zeta_2} \ , \eqno (2.34)$$
of which only that with the $+$ sign obeys the boundary condition at
$r = 0$.    When $\zeta_2 < 0$ a more detailed discussion is given
later.  If the series for $\eta$ does not terminate the asymptotic
form of the recurrence relation gives $a_p \simeq 2 a_{p-1}/p$ so
$\eta \propto e^{2 \tilde r}$ and $\psi_r$ is divergent at $\infty$;
so the series must terminate at $a_{P-1}$ say and in (2.32)
$$\tilde \zeta = 2 (P+s) + 3(N-2)=2(P+s+\beta ) \ , \eqno (2.35)$$ 
with $s$ given by taking the upper sign in (2.34) (i.e., $s = L$ when
$\zeta_2 = 0$).

Remembering that $\tilde \zeta = \zeta/\alpha$ and that $\alpha^2 = -2
ME \hbar ^{-2}$ expression (2.35) can be recast as the energy spectrum
$$E = - {\hbar ^2 \over 8M} \, {\zeta^2 \over (P + s + \beta)^2} = -
{\hbar^2 \over 2M} \, {\zeta^2 \over \left[2P-1 + \sqrt{ \left(2L
+3N-5 \right)^2 + 4 \zeta_2} \right]^2} \ . \eqno (2.36)$$ In
accordance with our theorem the energy levels with general $N$ are
given by putting $N=2$ and then replacing $L$ by $L+\beta = L + {3
\over 2} (N-2)$.  Of course if $\zeta_2=0$ we have $s=L$ and the
theorem is then obvious from the first form.  Notice that the theorem
really applies to $\alpha^2 = -2ME/\hbar^2$ thus we can only apply it
to $E$ itself if we consider a two body problem with the same mass $M$
as the N-body problem; $\zeta^2$ is also taken as unchanged since we
require both problems to have the same $U(r)$.  However this in no
way restricts us to N-body problems with $M$ and $\zeta$ independent
of $N$; it merely means that we change correspondingly the masses $M$
and coefficients $\zeta$ in the two body problems with which we
compare N-body problems of different $N$.

Some may wish to see the precise Schr\"odinger hydrogen atom spectrum
with the correct reduced mass emerging when $N=2$; to get this we must
evaluate $\zeta$ in terms of $Ze^2$.  Our potential energy is $V =
-{\textstyle {1 \over 2}} \hbar^2 M^{-1} \zeta/r$ but this $r$ is not
the separation of the nucleus from the electron, $R$, but the mass
weighted r.m.s. separation of them from the centre of mass.  Hence $r
= (mm_n/M^2)^{1/2} R$ where $m$ and $m_n$ are the masses of the
electron and the nucleus respectively.  Setting $V = - {Ze^2 \over R}$
we deduce that $\zeta = 2 (mm_n)^{1/2} \hbar ^{-2} Ze^2$.  Inserting
this $\zeta$ into (2.36) along with $N=2, \zeta_2 = 0, n = P+L, M =
m+m_n$ and putting the reduced mass $mm_n/M = m_r$ the energy levels
of hydrogenic atoms are given by
$$E = - {m_r \over 2 \hbar^2} \, {(Ze^2)^2 \over n^2} $$
just as they should be.

We now return to the question of how negative $\zeta_2$ can be.  Since
$L$ can be zero the energy of the ground state ceases to
be real if $\zeta_2  < - \left( {3N-5 \over 2} \right)^2$ which gives $\zeta_2 < - {1 \over 4}$ for the familiar
case $N=2$.  Such strongly attractive forces cause the particles to
propagate into the nucleus and the ground state ceases to exist.  It
may be seen that the limiting case has a wave function $\psi \propto
r^{- {1 \over 2}}$ near the origin which is easily still square
integrable $\int \psi \psi ^\star r^{2} dr < \infty$.  This is also
true for the limiting case $\zeta_2 = - \left( {3N-5 \over 2} \right) ^2$ for then $\psi
\propto r^{-{1 \over 2}(3N-5)}$ and $\int \psi \psi ^\star r^{3N-4} dr
< \infty$.  The limits are surpassed for the attractions of magnetic
monopoles on the magnetic moments of protons and for charged monopoles
attracting spinning electrons (Lynden-Bell \& Nouri-Zonoz 1998).

\section{Level Degeneracies}

In equations (2.26) and (2.27) and the attendant discussion we showed
that the solutions of our Schr\"odinger equation consisted of a
hyperspherical harmonic in $3N-3$ dimensions times a radial function.
Furthermore the hyperspherical harmonics of degree $L$ in $D$
dimensions are the coefficients of $r^L$ in the polynomial solutions
of Laplace's equation in $D$ dimensions.  Thus the degeneracy of the
states of given $L$ and given radial quantum number $P$ will be equal
to the number of independent polynomial solutions of Laplace's
equation of degree $L$ in $D=3N-3$ dimensions (i.e., harmonic
polynomials).  To determine this number we first ask how many
independent polynomials of degree $L$ exist in $D$ dimensions without
the harmonic requirement.  Each can be considered as a term of the
form $\Pi _i x_i^{l_i}$ where $i$ runs from 1 to $D$ and $\Sigma l_i
=L$.  That number of polynomials is equal to the number of ways of
dividing $L$ objects into $D$ groups where a group is allowed to
contain no objects.  If we take $L$ units and $D-1$ dividing bars then
the number of ways of ordering them is $(L+D -1)!$ and if we disregard
the ordering of the $D-1$ bars among themselves and the $L$ units
among themselves the number of ways of sorting them into groups is  
$$G (L,D) = {(L+D-1)! \over (D-1)!L!}\ , \eqno (3.1)$$ so this is the
number of independent polynomials of degree $L$.  Let $f_L ({\bf
x}_a)$ be such a homogeneous polynomial of degree $L$ in the ${\bf
x}_a$.  In general 
$$\nabla^2 f_L = \sum_a {\partial \over \partial {\bf x}_a} \cdot {\partial \over \partial {\bf x}_a} \, f_L = f_{L-2}$$ where $f_{L-2}$ is such a polynomial of degree $L-2$ which will have $G(L-2, D)$ independent coefficients.  The condition that $f_L$ be harmonic $(\nabla^2f_L = 0)$ thus imposes $G(L-2, D)$ constraints on the $G(L, D)$ free coefficients in $f_L$.
 Thus the number of independent harmonic polynomials of degree $L$
in $D$ dimensions is
$$\displaylines{\  g (L,D) = G(L,D) - G(L-2,D) = \hfill \cr \hfill{}\cr
\hfill \quad \ \, = {(L+D-3)! \left [ (L+D-1)(L+D-2)-L(L-1)\right] \over
(D-1)!L!} = \hfill (3.2) \cr \hfill{}\cr = {(L+D-3)! \over (D-2)!L!}
(2L+D-2) \ . \qquad \qquad \qquad \qquad \qquad \quad \ \ \,  }$$
Notice that for the familiar case $D=3$ this gives the correct answer,
$2L+1$, for the degeneracy of the states of given $L$.

When $\zeta_2=0$ we have the extra degeneracy between the $s, p, d, f$
levels of hydrogen.  Then a state of principal quantum number $n$ can
be obtained by combining a wave function of given $L$ with a radial
wave function with radial quantum number $P = n - L \geq 1$.  Thus, to
find the degeneracy of states with a given $n$, we need to know the
number of harmonic polynomials of degree less than or equal to $n-1$,
because the $n-L-1$ extra quanta are taken up by different radial wave
functions.  To find this number we merely add a dummy group `$o$' to
our sorting of $L$ objects into groups and ignore the number of units,
$n_o$, that falls into that group.  Thus the required answer is 
$$g_H (n,D) = g (n-1, D+1) = {(n+D-3)! \over (D-1)! (n-1)!} (2n + D-3) \ . 
\eqno (3.3)$$ 
For $D=3$ this reduces to $n^2$, which is the
familiar degeneracy of the $n^{\rm th}$ hydrogen level before spin is
considered.  Thus for our problems the degeneracy of levels of
hyper-angular-momentum $L$ for a system of $N$ particles is $g(L,
3N-3)$ with $g\ {\rm given}$ by (3.2), while for $\zeta_2=0$ the degeneracy of
the $n^{\rm th}$ level is \break $g_H (n, 3N-3) = g (n-1, 3N-2)$.

The above degeneracies are worked out assuming that none of the
particles are identical.  In practice we are more interested in
problems with $N$ identical Bosons or $N$ identical Fermions and they
require wave functions with even or odd permutational symmetries so we
now study that question.

\section{Symmetry under permutation of particles}

\subsection{Bosons}

For Bosons we need wave-functions that are symmetrical for the
interchange of any two particle labels ${\bf x}_I, {\bf x}_J$.  Both
${\overline {\bf x}}$ and $r$ have the required symmetry when the
particles are of equal mass.  Just as the magnitude of the angular
momentum treats $x, y$ and $z$ symmetrically in 3 dimensions, so the
magnitude of the hyperangular momentum $L$ is symmetrical for particle
interchange.  However the components of the term $L_{ij}$ and the
vector ${\hat {\bf r}}$ are not symmetrical under the interchange of
particle labels.  In \sect2 we found the solutions for our
$N$-particle wave functions in the form
$$\psi = \exp (i {\bf k} \cdot {\overline {\bf x}}) Y_L ({\hat {\bf
r}}) \psi _r (r) \ , \eqno (4.1)$$
\vfill
\eject

\noindent
where $\psi_r$ depends only on the scalar magnitude $L$.  The only
term that is not automatically symmetrical for particle interchange is
$Y_L ({\hat {\bf r}})$ but even that will be automatically symmetrical
when $L=0$, because $Y_0$ is constant.  Thus the ground state and all
$s$-states are automatically symmetrical and are possible states for a
system of identical Bosons.  

The states we have been discussing are {\em not} the one-particle
states commonly considered as components of $N$-particle product
states (or, for Fermions, Slater determinants); rather our states are
themselves $N$-particle states.  To get a symmetrical $N$-particle
state from one lacking that symmetry we merely add all the wave
functions obtained by permuting the labels on the particles.  But
whereas each of our wave-functions thereby generates one boson
$N$-particle state, such a state in general comes from a number of
different unsymmetrical wave-functions so we can no longer count the
degeneracies by the arguments of \sect3.  However the arguments of
\sect3 connect the number of $Y_L$ functions with the number of
polynomials that are homogeneous and both of degree $L$ and harmonic.
If we can count interchange symmetric polynomials independent of
${\overline {\bf x}}$ which are homogeneous of degree $L$ and
solutions of Laplace's equation, we have the degeneracy of the quantum
states of hyper-angular-momentum $L$.  

Let $F_L ({\bf x}) = F_L ({\bf x}_1, {\bf x}_2, \ldots {\bf x}_N)$ be a
homogeneous polynomial of degree $L$ in ${\bf x}$ which is symmetric
under the interchange of any ${\bf x}_I$ with ${\bf x}_J$.  Then \break $F_L
(\lambda {\bf x}_1, \lambda {\bf x}_2, \ldots \lambda {\bf x}_N) =
\lambda^L F_L ({\bf x}_1, {\bf x}_2, \ldots  {\bf x}_N) = \lambda^L
F_L ({\bf x}_1, {\bf x}_N, \ldots  {\bf x}_2), \ {\rm etc.} $ Consider
$F_L ({\bf x} - {\overline {\bf x}} ) = F_L ({\bf x}_1 - {\overline
{\bf x}}, \ldots  {\bf x}_N - {\overline {\bf x}})$.  It is also a
homogeneous polynomial of degree $L$ in ${\bf x}$ and is also
symmetric, but it has the property that it is invariant under the
transformation ${\bf x}_I \rightarrow {\bf x}_I + {\mbox {\boldmath
$\Delta$}}$ for all $I$ (because then ${\overline {\bf x}} \rightarrow
{\overline {\bf x}} + {\mbox {\boldmath $\Delta$}}$).  Thus such
functions do not depend on the position of the centre of mass.
However it can happen that $F_L ({\bf x} - {\overline {\bf x}})$ is
identically zero even when $F_L ({\bf x})$ is not.  For this to happen
$F_L ({\bf x})$ must be of the form $F_L ({\bf x}) = {\overline {\bf x}}
\cdot {\bf F}_{L-1} ({\bf x})$ where ${\bf F}_{L-1} ({\bf x})$ is a
vector each of whose components is a polynomial of degree $L-1$ in
${\bf x}$ which is symmetric under interchange of ${\bf x}_I$ and
${\bf x}_J$.  Now let ${\overline G}(L)$ be the number of independent
symmetric polynomials which are homogeneous of degree $L$ in 3$N$
dimensions.  Then the number of such polynomials giving rise to
non-zero $F_L ({\bf x} - {\overline {\bf x}})$ will be ${\overline G}
(L)$ less the number of free coefficients in the $- {\overline {\bf
x}} \cdot {\bf F}_{L-1}$ term which we might expect to be $3
{\overline G} (L-1)$.  However that is not quite right because a
polynomial with a factor ${\overline {x}}\, \overline {y}$ will occur
as a possibility in both the $x$ and $y$ components of ${\bf F}_{L-1}$
and in subtracting $3 {\overline G} (L-1)$ we will have subtracted its
number of free coefficients not once but twice.  The same double
counting will have occurred for polynomials with factors ${\overline
y} \, {\overline z}$ or ${\overline z} \, {\overline x}$ so we should
be subtracting not $3 {\overline G} (L-1)$ but rather $3 {\overline G}
(L-1) - 3 {\overline G} (L-2)$.  However even that is not quite
correct because there may be polynomials with a factor ${\overline
x}\, {\overline y}\, {\overline z}$.  They will have been subtracted
off three times in $3 {\overline G} (L-1)$ but added back in three
times in $3 {\overline G} (L-2)$ so they will still be there and they
should not be since they clearly vanish when ${\bf x} - {\overline
{\bf x}}$ is written for ${\bf x}$.  Thus finally the number of
independent symmetric polynomials which are homogeneous of degree $L$
and independent of ${\overline {\bf x}}$ is 
$${\overline G}_1 (L) \equiv {\overline G} (L) - 3 {\overline G} (L-1)
+ 3 {\overline G} (L-2) - {\overline G} (L-3) \ . \eqno (4.2)$$
However, we still have to impose Laplace's equation
$$\sum_I \nabla^2_I F_L \equiv \nabla^2 F_L=0 \ . $$ Now in general
$\nabla^2 F_L ({\bf x} - {\overline {\bf x}})$ will be a polynomial of
degree $L-2$ in ${\bf x}$.  However, since $F_L ({\bf x} - {\overline
{\bf x}})$ is invariant to the transformation ${\bf x} \rightarrow
{\bf x} + \mbox {\boldmath $\nabla$}$, $\nabla^2 F_L ({\bf x} -
{\overline {\bf x}})$ will also have that property.  Thus in general
we may write $$\nabla^2 F_L ({\bf x} - {\overline {\bf x}}) = F_{L-2}
({\bf x} - {\overline {\bf x}})$$ where $F_{L-2}$ is also symmetrical
for particle label interchange since $\nabla^2$ does not destroy that
property.  Thus the condition $F_{L-2} ({\bf x} - {\overline {\bf x}})
\equiv 0$ will put ${\overline G}_1 (L-2)$ constraints on the
${\overline G}_1 (L)$ free coefficients of the homogeneous $L^{{\rm
th}}$ degree polynomial $F_L ({\bf x} - {\overline {\bf x}})$.  There
will be just ${\overline G}_1 (L) - {\overline G}_1 (L-2)$ free
coefficients left in $F_L ({\bf x} - {\overline {\bf x}})$ after
imposing the harmonic condition so this is the degeneracy of the $Y_L$
that corresponds to the $(2L+1)$ with $L$ even for the 2 boson
problem.  Since ${\overline G}_1$ is known in terms of ${\overline G}$,
we have reduced our problem to that of determining the number of
exchange-symmetric homogeneous polynomials of degree $L$ in 3$N$
dimensions.  This is the crux of our problem and it took us
considerable thought to solve it.  Exchange symmetry
involves exchanging vectors ${\bf x}_I$ with ${\bf x}_J$, so we do not
need symmetry in all 3$N$ dimensions but only between them taken in
triples.  We shall begin our
considerations with the simpler case of $N$ bosons on a line with each
having but one coordinate  $x_I$.  We then wish to know how many
independent exchange-symmetric polynomials there are which are
homogeneous of degree $L$ in $N$ dimensions.

Let $\Phi = S \Pi_I x^{\ell _I}_I$ be a symmetrical polynomial of
degree $L$ with $N$ factors $x^{\ell _I}_I$.  $S$ is the symmetrising
operator which is the sum over all permutations of the particle labels
$I$.  Different symmetric polynomials are characterised by different
sets of integers $(\ell_1, \ell_2, \ldots \ell_N)$ or partitions of
the integer $L$ into $N$ parts, some of which may be zero.  We
construct the generating function
$$B_1 (u,x) = \sum^\infty_{n=1} \sum^\infty_{\ell = 0} p(n, \ell) u^n
x^\ell$$
where $p(N,L)$ is the number of partitions of $L$ into $N$ integers
that may be zero, and for convenience we have defined $p (0, \ell) =
0$ and $p (n, 0 ) = 1$.  We now show how the theory of partitions
allows us to construct the function $B_1 (u, x)$.

\subsection{Partitions of an integer $L$}

We learn from Abramowitz \& Stegun (1964) that `The number of
decompositions of an integer $L$ into integer summands without regard
to order is called $p (L)$'.

For example, five may be written
$$5 = 4+1 = 3+2 = 3+1+1 = 2+2+1 = 2+1+1+1 = 1+1+1+1+1 \ , \eqno
(4.3)$$
so we deduce that $p(5) =7$.  It is easiest to work with the
generating function for the $p(L)$ which we call $A(x)$.  For this
there is a standard result see, e.g., Hardy \& Ramanujan (1918),
$$A(x) = 1 + \sum^\infty_1 p(L) x^L = \prod\limits^\infty_{\ell =1} (1 -
x^\ell)^{-1} \ . \eqno (4.4)$$
\vfill
\eject

For what follows it is essential to understand how this standard
result comes about.  To do so we rewrite the product by expansions in
powers of $x$
$$\begin{array}{lccc}
A(x) = & \! \! \left( 1 + x^1 + x^2 + \ldots \right) & \! \! \left( 1 + (x^2)^1 +
(x^2)^2 + \ldots \right) & \! \! \left( 1 + (x^3)^1 + (x^3)^2 \ldots
\right) \cr
&({\rm generates \ units}) & ({\rm generates \ twos}) & ({\rm generates
\ threes })
\end{array}$$
\vskip -0.3cm
\rightline {(4.5)}
\vskip -0.6cm
$$\hfill \begin{array}{ccc}
\left( 1+ (x^4)^1 + \ldots \right) & \left( 1 + (x^5)^1 + \ldots \right)
&(\ldots) \cr
({\rm generates \ fours}) & ({\rm generates \ fives})& \qquad \qquad \qquad .
\end{array}$$ 
To see how the coefficient of $x^5$ in this expression is $p(5)=7$ we
first notice that we must take the 1 from all brackets after the fifth,
since otherwise we would get too high a power of $x$.  In the 5$^{\rm
th}$ bracket we can take the $x^5$ term but then we must take the 1
from all earlier brackets.  Alternatively we take the 1 in the 5$^{\rm
th}$ bracket.  In the latter case we turn to the fourth bracket.  Here
we may take the $x^4$ term but that can only be coupled to the $x$
term in the first bracket in which case we get the split of 5 into
$4+1$.  Turning now to the third bracket and taking the $x^3$ term we
can take it with either the $(x^2)^1$ bracket of the second term to
yield $3+2$ or with the $(x^1)^2$ term of the first bracket to yield
$3+1+1$.  Similarly from the second bracket we could take the
$(x^2)^2$ term with the $x$ from the first bracket to give $2+2+1$ or
the $(x^2)^1$ term with the $x^3$ from the first to give $2+1+1+1$.
Finally we could take the $x^5$ from the first bracket to give
$1+1+1+1+1$.  Thus the first bracket yields the number of ones in the
sum, the second the number of twos, the third the number of threes,
etc., and in this way the coefficient of $x^L$ yields $p(L)$.  

However, we need the restricted partition of $L$ into $N$ or fewer
non-zero integers $p(N,L)$.  These are sums of partitions $p_1(N,L)$
into exactly $N$ non-zero integers.  Looking at (4.3) we see, for
example, that $p_1(2,5) = 2 = p_1 (3,5)$.  If we place a factor $u$
along with each factor $x^{\ell}$ in (4.5), then the power of $u$ in
each term will tell us how many parts there are in the partition
generated by a particular term.  Thus in place of $A(x)$ we consider
$$\begin{array}{lcc}
A(u, x) = & \left(1 + ux + (ux)^2 + (ux)^3 + \ldots
\right) & \left( 1 + (ux^2)^1 + (ux^2)^2 + \ldots \right) \cr
 & ({\rm generates \ units}) & ({\rm generates \ twos}) \cr
&&\hfill (4.6)\cr
&\left( 1 +
(ux^3)^1 + (ux^3)^2 + \ldots \right)  \ \ \ \ldots \cr  
&({\rm generates \ threes}) \ \ \ \
\end{array}$$
Then the terms in $u^Nx^L$ will have exactly $N$ integers in the
corresponding partition of $L$ so (cf. 4.4)
$$A(u,x) = 1 + \sum^\infty_{n=1} \sum^\infty_{\ell-1} p_1(n,\ell)
u^nx^\ell = \prod\limits^\infty_{\ell =1} (1 - ux^\ell)^{-1} \ . $$
However, we want the number of partitions with $N$ or fewer integers,
i.e., \break $p(N,L) = \sum^N_{n=1} p_1 (n,L)$.  These sums will be
automatically generated if we multiply $A$ by $(1 + u + u^2 + u^3
\ldots)$ before taking the coefficient of $u^N x^L$ so 
$$B_1 (u,x) = \sum^\infty_{n=1} \sum^\infty_{\ell = 0} p(n,\ell) u^n
x^\ell = (1 + u +u^2 + \ldots) A(u,x) = \prod\limits ^\infty _{\ell = 0} (1 -
ux^\ell)^{-1} \ . \eqno (4.7)$$
\vfill
\eject

Readers will recognise the analogy of this expression with the grand
partition function for a gas of non-interacting Bosons.  So $p(N,L)$
can be found from the product as the coefficient of $u^Nx^L$.  Had we
been interested in Bosons on a line then $p(N,L)$ would have given us
the desired function ${\overline G} (L)$ but our problem is three
dimensional.  Instead of partitioning $L$ into integers $\ell_I$ we
need it partitioned into integer triples $\left( \ell_{Ix}, \ell_{Iy},
\ell_{Iz}\right) $ and the general term in our polynomial will be
$x^{\ell_{1x}}_1 y_1^{\ell_{1y}} z_1^{\ell_{1z}} x_2^{\ell_{2x}}
\ldots x_N^{\ell_{Nx}} y_N^{\ell_{Ny}} z_N^{\ell_{Nz}}$.  When we
permute we do so by exchanging $(x_1, y_1, z_1)$ as a triple with say
$(x_2, y_2, z_2)$.  The degree of our polynomial is 
$$L= \sum^N_{I=1} \left( \ell_{Ix} + \ell _{Iy} + \ell_{Iz} \right) \
. $$
A triple $(2,1,0)$ corresponding to $\ell_{1x} = 2, \ell_{1y} = 1,
\ell_{1z} = 0$ will not be permuted into $(1,0,2)$ by exchanging
particle labels so such triples must be regarded as distinct.  The
number of partitions of 3 into triples, such that the order within a
triple matters but the order of the different triples does not,
$p_3(L)$ with $L=3$, is already quite a handful.  Writing $=$ for an
equal total we have 
$$\left( \begin{array}{c}
3 \\
0 \\
0 \end{array}
\right) =
\left( \begin{array}{c}
0 \\
3 \\
0 \end{array}
\right)
=
\left( \begin{array}{c}
0 \\
0 \\
3 \end{array}
\right)
=
\left( \begin{array}{c}
2 \\
1 \\
0 \end{array}
\right)
=
\left( \begin{array}{c}
0 \\
2 \\
1 \end{array}
\right)
=
\left( \begin{array}{c}
1 \\
0 \\
2 \end{array}
\right)
=
\left( \begin{array}{c}
1 \\
2 \\
0 \end{array}
\right)
=
\left( \begin{array}{c}
0 \\
1 \\
2 \end{array}
\right)
=$$
$$
=
\left( \begin{array}{c}
2 \\
0 \\
1 \end{array}
\right)
=
\left( \begin{array}{c}
1 \\
1 \\
1 \end{array}
\right)
=
\left( \begin{array}{c}
2 \\
0 \\
0 \end{array}
\right)
+
\left( \begin{array}{c}
1 \\
0 \\
0 \end{array}
\right)
\ =
\begin{array}{ll}
& {\rm 8 \ more \ like \ that} \cr 
&{\rm with} \ x,y,z \ {\rm components} \cr
& {\rm permuted.} 
\end{array}
\ =
$$
$$
\hskip 3.4cm
\ =
\left( \begin{array}{c}
1 \\
1 \\
0 \end{array}
\right)
+
\left( \begin{array}{c}
1 \\
0 \\
0 \end{array}
\right)
\! =
\begin{array}{ll}
& {\rm 8 \ more \ like \ that} \cr 
&{\rm with} \ x,y,z \ {\rm components} \cr
& {\rm permuted \ in \ each \ triple.} 
\end{array}
\ =
$$
$$
\hskip 1cm
=
\left( \begin{array}{c}
1 \\
0 \\
0 \end{array}
\right)
+ 
\left( \begin{array}{c}
1 \\
0 \\
0 \end{array}
\right)
+
\left( \begin{array}{c}
1 \\
0 \\
0 \end{array}
\right)
=
\begin{array}{ll}
& {\rm 9 \ more \ like \ that} \cr 
&{\rm (not \ 26 \ because \ the}\cr
& {\rm ordering \ of \ the \ triples} \cr
&{\rm  does \ not \ count).} 
\end{array}
$$
Hence there are 38 partitions of 3 into triples!  Now let $p_3 (L)$ be
the number of partitions of $L$ into triples, every triple being
counted as different but the different orderings of the same triples
being regarded as the same.  By analogy with (4.5) we consider the
expression

\eject

$$
A(x,y,z)\equiv $$
$${\setlength{\arraycolsep}{.1em}
\begin{array}{lllll}
\equiv (1 + x + x^2 ...)&         \times &         (1 +
(x^2)^1 + (x^2)^2 + ...)&         \times &         (1 +
(x^3)^1 + (x^3)^2 + ... )\times ... \cr
\times \, (1 + y + y^2 ...) &         \times &         (1 +
(y^2)^1 + (y^2)^2 + ...)&         \times &         (1 +
(y^3)^1 + (y^3)^2 + ... ) \cr
\times \, (1 + z+ z^2 ...) &        \times &         (1 +
(z^2)^1 + (z^2)^2 + ...) &         \times &        (1 +
(z^3)^1 + (z^3)^2 + ... ) \cr
&         \times &         (1 + (yz)^1 + (yz)^2 + ...) &    
    \times &         (1 + (x^2y)^1 + (x^2y)^2 ...)\cr
{\rm This \ column} &         \times &         (1 + (zx)^1 +
(zx)^2 + ...) &         \times &         (1+ (y^2z)^1 +
(y^2z)^2 + ...) \cr
{\rm generates} &         \times &         (1+(xy)^1 +(xy)^2 +
...) &         \times &          (1+ (z^2x) + ...)\cr
{\rm triples} & &     {\rm column \ generates} &          \times &  
       (1 + (x^2z) + ...)\cr

\left( \!   \begin{array}{c}
1 \\
0 \\
0  \end{array} 
\!     \right) 
 \!    {\rm or} \!    
\left(\!      \begin{array}{c}
0 \\
1 \\
0  \end{array} \!     \right) 
\!     {\rm or}  \!   
\left(\!      \begin{array}{c}
0 \\
0 \\
1  \end{array}\!      \right)&&
\hskip -.1cm 
 \left(\!      \begin{array}{c}
2 \\
0 \\
0  \end{array} 
\!     \right) \! \!    
\left(  \!    \begin{array}{c}
0 \\
2 \\
0  \end{array}  \!   \right)  \! \!   
\left(  \!    \begin{array}{c}
0 \\
0 \\
2  \end{array}\!     \right) \! \!   
\left(  \!    \begin{array}{c}
1 \\
1 \\
0  \end{array} 
 \!    \right) \! \!    
\left(  \!    \begin{array}{c}
0 \\
1 \\
1  \end{array}\!      \right)  \! \!   
\left( \!     \begin{array}{c}
1 \\
0 \\
1  \end{array} \!     \right)&&
\hskip -0.45cm
\begin{array}{rl}
\times         &(1 + (y^2x) + ...)\cr
\times         & (1 + (z^2y) + ...)\cr
\times         & (1 + (xyz) + (xyz)^2 + ...)
\end{array}\cr

{\rm and \ multiple}&& {\rm and \ multiple} && {\rm generates \
triples} \cr
{\rm combinations} && {\rm combinations} && {\rm with}\ \ell_x +
\ell_y + \ell_z =3 \cr
{\rm thereof} && {\rm thereof} && 
\end{array}}
$$

By analogy with the arguments beneath (4.5) one sees how the terms of
the third degree generate all the partitions of 3 into triples that we
have just enumerated.  So the coefficient of $t^L$ in $A(t,t,t)$ will
be $p_3(L)$, the number of partitions of $L$ into triples.
Furthermore $A(x,y,z)$ may be compactly written in the form
$${\setlength{\arraycolsep}{.1em}
\begin{array}{lcl}
A(x,y,z) = & \prod\limits^\infty_{p=0}\prod\limits^\infty_{q=0}
\prod\limits^\infty_{r=0} & (1-x^py^qz^r)^{-1} \ . \cr
& ^{p + q + r \neq 0} &\cr
\end{array}}
$$
\vskip -1.7cm
$$\eqno{(4.8)}$$
\vskip .8cm
For our $N$ boson problem we are interested not in such partitions
into any triples but in partitions constrained to have $N$ or fewer
triples as summands.  To get exactly $N$ summands we merely insert a
$u$ in each term as was done in (4.6).  While to get the sum of all
terms with $N$ or less summands we have to multiply by $(1+u+u^2
\ldots)$ as in (4.7).  Thus putting $x=y=z=t$ the Boson generating
function in 3 dimensions is, writing $\ell = p+q+r$,
$$B(u,t) = \prod\limits^\infty_{p=0} \prod\limits^\infty _{q=0}
\prod\limits^\infty _{r=0} \left(1 - ut^{p+q+r} \right)^{-1} =
\prod\limits^\infty_{\ell = 0} \left(1 - ut^\ell \right)^{-{1 \over 2} (\ell
+1)(\ell +2)} \, \eqno (4.9)$$
where ${1 \over 2} (\ell +1)(\ell +2)$ is the number of terms in the
triple product with $p+q+r = \ell$.  

The required function ${\overline G} (L)$ is the coefficient of $u^N
t^L$ in $B(u,t)$.  Again if one only wants ${\overline G} (L)$ for
$L<L_{\max}$ then the infinite product can be replaced by a finite
product up to $L_{\max}$ without altering the required coefficients.

To get ${\overline G}_1(L)$ one merely takes the coefficient of
$u^Nt^L$ in $(1 - t)^3B(u,t)$ \break while ${\overline G}_1 (L) - {\overline
G}_1 (L-2)$ is the coefficient of $u^Nt^L$ in $(1 -t^2) (1 -t)^3
B(u,t) = (1+t) (1-t)^4 B(u,t)$.

In summary the degeneracy $g_B(L)$ of the $N$ Boson state with
hyperangular momentum $L$ is the coefficient of $u^N t^L$ in the
expression
$$(1+t) (1-t)^4 \prod\limits^\infty_{n=0} \left( 1 -
ut^{\ell}\right)^{-{1 \over 2} \left(\ell + 1 \right) \left( \ell +2
\right)} \ \ . \eqno (4.10)$$
For two particles $N=2$ we find that the coefficient of $u^2$ is
$\left( 1-3t^2 \right) \left( 1 - t^2 \right)^{-2}$.  The coefficient
of $t^L$ in this expression is $2L+1$ when $L$ is even and zero when
$L$ is odd just as it should be.  This is just as in the $C^{12} - C^{12}$
homonuclear diatomic molecule of two Bosons with every odd rotational state
missing as seen in Carbon star spectra.  

\subsection{Fermions}

The argument of \sect3 relates the number of hyperspherical
harmonics of degree $L$ to the number of harmonic polynomials of that
degree.  Our experience in \sect4$\, a$ leads us to study first the
number of antisymmetric polynomials of degree $L$ in $N$ dimensions.

If $x^{\ell_1}_1 x^{\ell_2}_2 \ldots x^{\ell_N}_N$ is a term in such a
polynomial then $\Sigma \ell_I = L$.  Furthermore the
antisymmetric polynomial involving that term is the Slater determinant
$$\left| \begin{array}{ccccl}
x^{\ell_1}_1 & x_2^{\ell_1} & \ldots & x^{\ell_1}_N \cr
x^{\ell_2}_1 & x^{\ell_2}_2 & \ldots & x^{\ell_2}_N \cr
\vdots &&&\vdots \cr
x^{\ell_N}_1 & x^{\ell_N}_2 & \ldots & x^{\ell_N}_N
\end{array}
\right| 
$$
Clearly if $\ell_I = \ell_J$ for $I \neq J$ then this determinant
vanishes.  Furthermore the determinant only changes sign (at most) if
the $\ell_I$ are permuted.  So for any term that survives we may,
without loss of generality, take $\ell_1 < \ell_2 < \ell_3 \ldots <
\ell_N$.  Thus among the $\ell_I$ only $\ell_1$ can be zero,  $\ell_2$ must
be at least 1, $\ell_3$ at least 2 and so on with $\ell_N$ at least $N-1$.
For a non-zero result $L= \Sigma \ell_I \geq {\textstyle {1 \over
2}} (N-1)N$.

By analogy with our study of partitions $p (L)$ for the Boson case we
now study partitions of $L$ into distinct parts.  Let $q(L)$ be the number
of decompositions of $L$ into distinct integer summands without regard
to order.  Thus $5 = 4+1 = 3+2$ so that $q(5) = 3$.  The generating
functions for the $q(L)$ is, setting $q(0)=1$,
$$\sum^\infty _0 q(L) x^L = \prod\limits^\infty_{\ell = 1} (1 + x^\ell) \
. \eqno (4.11)$$
However not all decompositions of $L$ into distinct parts lead to
antisymmetric polynomials in $N$ dimensions.  We need a decomposition
into either $0 + (N-1)$ unequal integers or into $N$ unequal integers,
i.e., we need the coefficient of $u^N x^L$ in
$$\prod\limits^\infty_{\ell =0} \left( 1 + ux^\ell \right) =
\sum^\infty_{L=0} \sum^\infty _{N=0} q (L,N) u^N x^L \ , \eqno
(4.12)$$
which is the expression analogous to (4.7) of the Boson case.  Again
it is the analogue of the grand partition function for Fermions. 

The generalisation corresponding to three dimensions follows the
argument for (4.8) and yields the generating function
$${\setlength{\arraycolsep}{.1em}
\begin{array}{lcl}
E(x,y,z) = & \prod\limits^\infty_{p=0}\prod\limits^\infty_{q=0}
\prod\limits^\infty_{r=0} & (1+x^py^qz^r) \ , \cr
& ^{p + q + r \neq 0} &\cr
\end{array}}
$$
\vskip -1.8cm
$$\eqno{(4.13)}$$
\vskip .7cm
\noindent
which leads analogously to (4.9) to the Fermion generating function
$$F(u,t) = \prod\limits^\infty_{p=0} \prod\limits^\infty_{q=0}
\prod\limits^\infty_{r=0} \, \left( 1 + u t^{p+q+r} \right) =
\prod\limits^\infty _{\ell = 0} \left( 1 + ut^{\ell}\right) ^{{1 \over
2} (\ell +1)(\ell +2)} \ . \eqno (4.14)$$
The coefficient of $u^Nt^L$ in this expression gives the number of
homogeneous polynomials of degree $L$ antisymmetric for interchanges
of triples in 3$N$ dimensions.  For Fermions of spin ${\textstyle {1
\over 2}}$ we do not need complete antisymmetry but only antisymmetry
between particles of the same spin state.  To allow for the $\alpha$
and $\beta$ spin states being alternatives we generalise $E (x,y,z)$
to
$$\prod\limits^\infty_{p=0} \prod\limits^\infty_{q=0}
\prod\limits^\infty_{r=0} \, \left( 1 + x^py^qz^r \alpha \right)
\left( 1 + x^py^q z^r \beta \right) \eqno (4.15)$$
and in place of $F(u,t)$ we then find $F^2$
$$F^2 = \prod\limits^\infty_{\ell = 0} \left( 1 + u t^\ell
\right)^{(\ell + 1)(\ell+2)} \ . \eqno (4.16)$$
The coefficient of $u^Nt^L$ in $F^2$ is the number of Fermionic
polynomials for spin ${1 \over 2}$ particles with the correct
antisymmetry.  The arguments relating the degeneracy of the
$N$-Fermion spin ${1 \over 2}$ wave function for the state of
hyper-angular momentum $L$ to this expression is the same as for the
Boson case.  Thus $g_F(L)$ is coefficient of $u^Nt^L$ in the expression
$$(1+t) (1-t)^4 \prod\limits^\infty_{\ell = 0} \left( 1 + ut^\ell
\right)^{(\ell +1) (\ell +2)} \ . \eqno (4.17)$$
To check this formula we take the coefficient of $u^2$, and find
for the 2 Fermion problem the generating function
$$\left( 1 + 9t + 3t^2 + 3t^3 \right) \left( 1 - t^2 \right) ^{-2} =
\sum_{\ell {\rm even}} (2 \ell + 1 ) t^\ell + 3 \sum_{\ell {\rm odd}} (2
\ell + 1 ) t^{\ell} \ , \eqno (4.18)$$
in which the coefficients of $t^\ell$ will be recognised as the
degeneracies of the rotational states of the hydrogen molecule with 2
protons of spin $1 \over 2$.  The even $\ell$ values correspond to
para-hydrogen and the odd $\ell$ values to ortho-hydrogen.

We now turn to the degeneracies in hydrogen-like potentials in which
different $L$ states can be degenerate.  Here we need the sum of the
$g_F(L)$ for all \break $L  \leq n-1$ as in equation (3.3).  If we multiply
our generating function for $g_F(L)$ by $1+t+t^2+t^3+ \ldots =
1/(1-t)$ and then pick the coefficient
of $u^Nt^{n-1}$ we will get the required sum so the hydrogenic Fermi
degeneracy is that coefficient in $$(1+t) (1-t)^3
\prod\limits^\infty_{\ell = 0} (1 + ut^\ell)^{(\ell + 1)(\ell + 2)}\
. \eqno (4.19)$$
To find the ground state we need the first energy level $n$ for which
the coefficient of $u^Nt^{n-1}$ is non-zero.  To get $u^N$ and no more
with the lowest power of $t$, we need to use the $ut^\ell$ terms
rather than the 1 in all the low $\ell$ brackets since others come
with higher powers of $t$.  Thus if the highest $\ell$ needed is
$\ell_m$ we require
$$\sum ^{\ell_m}_{\ell = 0} (\ell +1) (\ell +2) = N\ . \eqno (4.20)$$ The sum is
$\, {1 \over 3} (\ell _m +1)\, (\ell_m +2)\, (\ell _m + 3)\, $ but $\, N\, $ may not
be of precisely this form for  integer $\ell_m$, in which case we take
the lowest $\ell_m$ that gives \break  ${1 \over 3} (\ell _m +1) (\ell_m +2)
(\ell _m + 3) \geq N$ so that there will be at least one term in
$u^N$.  We are interested in the least power of $t$ associated with
this term. This will be $$n-1 = \sum^{\ell_m}_{\ell =1} \ell (\ell +1)
(\ell +2) = {{\textstyle {1 \over 4}}} \ell _m (\ell _m+1) (\ell_m +2)
(\ell _m + 3) \eqno (4.21)$$ whenever $N$ is of the special closed shell form given
by the equality.  Then \break $\ell_m \simeq (3N)^{1/3} - 2 - (3N)^{-1/3}$
and $n = {3N \over 4} \ell_m + 1 \rightarrow (3N)^{4/3}\big / 4$ as
$N \rightarrow \infty$.  For large $N$ we then find that the ground
state energy behaves as 
$$E = - {\hbar^2 \over 8M} \, {\zeta^2 \over \left[ {(3N) \over 4}
^{4/3} \right]^2} \ \, . \eqno (4.22) $$ To compare this energy with that of a white dwarf
star we must first choose an appropriate value of $\zeta$ so that the
potential corresponds to gravity.  We showed in an earlier paper
(Lynden-Bell \& Lynden-Bell 1999) that
at high temperatures our systems have a Gaussian density profile at
equilibrium.  For such a profile the potential energy may be expressed
in terms of the total mass $Nm_H$ and the rms radius at equilibrium
$r$ and is $$V = \left(  {3 \over 4 \pi}\right)^{1/2} G (Nm_H)^2/r\
. \eqno (4.23) $$  We therefore choose $${\hbar ^2 \over 2 M} \zeta = \left( {3
\over 4 \pi} \right) ^{1/2} G (Nm_H)^2 \ . \eqno (4.24)$$  With this choice our
ground state energy level is $$E = - {2 \ G^2m^4_H m_e \over 3^{5/3}\ 
\pi \ \hbar ^2} N^{7/3} \eqno (4.25)$$ where we have written $M = Nm_e$ for the
mass of the degenerate electrons whose wave function we have been
evaluating.  This expression is of precisely the form found by the
standard equation of state of a cold degenerate non-relativistic gas
under its own gravity.  We have, therefore, established that White
Dwarfs may be regarded as `superatoms' -- systems with N-body wave
functions which are solutions of Schr\"odinger's equation in a central
potential.  

\section{Conclusions}

By treating the N-body problem as a separable system in $3N$
dimensions we have shown that it can be solved in appropriate
potentials.  Marshall \& Wojciechowski (1988) have given the general
form of potentials that allow separability in many dimensions and we
have concentrated on the hyper-spherical one.  Of the many others
possible, a sub-class are symmetrical for exchange of the particles.

Whereas we have shown how to classify the wave functions by
hyper-angular momentum, those interested in rotating systems will need
to develop methods of classification involving the 3 dimensional
angular momentum; here the methods of Dragt (1965) and the work by
Louck \& Galbraith (1972) may prove useful.  It is hoped that study of
that problem will throw light on Bose-Einstein condensation of small
clusters in rotating systems.  

The fact that we are only able to treat non-relativistically
degenerate white dwarfs suggests that an appropriate generalisation
for relativistically moving particles should be sought for systems
that do not radiate gravitational waves.

In our earlier paper on the classical mechanics of these systems we
showed that the fundamental breathing oscillation in $r$ separates
for the far-more-general potentials $V= V_0 (r) + r^{-2} V_2 ({\hat
{\bf r}})$ where $V_2$ is an arbitrary function of all the coordinates
which is independent of scale, $\lambda$, when all the $x_I
\rightarrow \lambda x_I$.  While this is still the case in quantum
mechanics the energy eigenvalues depend on the other motions so this
separation does not by itself yield eigenvalues.  However, the
possibility of solid-like and liquid-like states where $V_0 = {1 \over
2} kr^2$ and \break $r^{-2} V_2 = {\sum \atop I} {\sum \atop {\! \! \! \! \!
<J}} k' \left| {\bf r}_I - {\bf r}_J \right|^{-2}$ suggests that such
systems are worthy of further study.

The statistical mechanics of the systems with $V = -{GM^2 \over r} \,
, r<r_e$, gives negative specific heats just as those studied earlier
as examples of phase transitions (Lynden-Bell \& Lynden-Bell 1977).
However, within one system there is no gravothermal catastrophe
(Antonov 1962, Lynden-Bell \& Wood 1968).

\begin{acknowledgments}
We thank Dr Dragt for sending us his papers relevant to classifying
our states by symmetry and Drs Twambey and Sridhar for referring  us
to the Calogero model.  D. Lynden-Bell is currently supported on a
PPARC Senior Research Fellowship.

We thank the referees for detailed comments that broadened the background of the paper and improved its clarity.
\end{acknowledgments}

\section*{Appendix}

\subsection*{Separation of $\nabla^2$ and commuting hyper-angular-momenta in $D$
dimensions}

We write $q_1 = r \cos \theta_1$, $q_2 = r \sin \theta_1
\cos \theta_2$, $q_a = r \sin \theta_1 \sin \theta_2 \ldots \sin
\theta_{a-1} \cos \theta_a $, \ldots , $q_{D-1} = r \sin \theta_1 \sin
\theta_2 \ldots \sin \theta_{D-2} \cos \phi$ and $q_D = q_{D-1} \tan
\phi$.

The metric is given by
$$\begin{array}{ll}
ds^2 &= dq^2_1 + dq^2_2 + \ldots dq^2_D =  \\
&  = dr^2 + r^2 \left( d \theta^2_1 + \sin^2 \theta_1 d\theta^2_2 + \sin^2
\theta^2_1 \sin^2 \theta^2_2 d\theta^2_3 + \ldots + {\displaystyle{\prod\limits^{D-2}_{i=1}}} \sin^2
\theta^2_i d \phi^2 \right) \\
& = dr^2 + r^2 \left( h^2_1 d \theta^2_1 + h^2_2 d \theta^2_2 + \ldots
+ h^2_{D-1} d\phi^2 \right) \ . \end{array}$$
$$\begin{array}{ll}
\nabla^2 & = r^{-(D-1)} {{\displaystyle{\partial \over \partial r}}} \left(r^{D-1}
{\displaystyle{{\partial \over \partial r}}} \right) + r^{-2} \left\{ \left
( {\displaystyle{\prod\limits^{D-1}_{i=1}}} h_i \right)^{-1} {\displaystyle{\sum^{D-1}_{j=1}}} {\displaystyle{{\partial \over
\partial \theta_j}}} \left[ \left( {\displaystyle{\prod\limits^{D-1}_{k=1}}} h_k \right)
h^{-2}_j {\displaystyle{{\partial \over \partial \theta_j}}} \right] \right\} \\
&= r^{-(D-1)} {\displaystyle{{\partial \over \partial r}}} \left(r^{D-1} {\displaystyle{{\partial
\over \partial r}}} \right) - r^{-2} A \ . \hfill ({\rm A}1) \end{array}$$
Since the $h_j$ depend only on $\theta_i$ with $i<j$ we can rewrite
the angular operator $A$ as follows
$$A = \sum^{D-2}_{j=1} h^{-2}_j \left( \sin \theta_j^{-D+1+j}\right)
{\partial \over \partial \theta_j} \left[ \left( \sin \theta_j
^{D-1-j} \right) {\partial \over \partial \theta_j} \right] +
h^{-2}_{D-1} {\partial^2 \over \partial \phi^2} \ . \eqno ({\rm A}2)$$

Using (2.27), Schr\"odinger's equation (2.25) separates giving in $D$
dimensions
$$L (L+D -2) = {\hat \psi}^{-1} A ({\hat \psi})\eqno ({\rm A}3)$$
multiplying (A2) or (A3) by $h^2_{D-1}$, $\hat \psi$ separates into
$\psi_\theta ({\mbox {\boldmath $\theta$}}) \psi_\phi (\phi)$ and the
separated equation for $\psi_\phi$ gives $\psi_\phi \propto \exp (im
\phi)$.  If we instead multiply (A3) by $h^2_{D-2}$ then the last two
terms are the only ones dependent on $\theta_{D-2}$ so the system
again separates and the $\theta_{D-2}$ equation is Legendre's equation
$${1 \over \sin \theta_{D-2}} \, {\partial \over \partial
\theta_{D-2}} \left( \sin \theta_{D-2} {\partial \psi_{D-2} \over \partial
\theta_{D-2}}  \right) - {m^2 \psi_{D-2} \over \sin^2
\theta_{D-2}} = - \ell_1 (\ell_1 +1) \psi_{D-2} \eqno ({\rm A}4)$$
with $\ell_1 > |m|$.

Similarly multiplying (A2) by $h^2_{D-3}$ the $\theta_{D-3}$ equation
is
$${1 \over \sin^2 \theta_{D-3}} \, {\partial \over \partial
\theta_{D-3}} \left( \sin^2 \theta_{D-3} {\partial \psi_{D-3} \over \partial
\theta_{D-3}} \right) - {\ell_1 (\ell_1 +1) \psi_{D-3} \over \sin^2
\theta_{D-3}} = - \ell_2 (\ell_2 +2) \psi_{D-3} \, \eqno ({\rm A}5)$$
where the eigenvalue on the right comes from (A3) applied in 4
dimensions and $\ell _2 \geq \ell_1$.

The general equation is
$$\displaylines{{1 \over \sin^{D-1-j} \theta_j} \, {\partial \over \partial
\theta_{j}} \left[ \sin^{D-1-j} \theta_{j} {\partial \psi_{D-1-j} \over \partial
\theta_{j}} \right] - {\ell_{j-1} (\ell_{j-1} +j -1) \psi_{D-1-j} \over \sin^2
\theta_{j}} \cr \hfill =  - \ell_j (\ell_j +j) \psi_{D-1-j} \, \hfill ({\rm A}6)}$$
where the eigenvalues on the right come from (A3) applied in $j+2$
dimensions. 

The operators on the left of (A4), (A5) and (A6) all have simultaneous
eigenvalues of the form $-\ell_j (\ell_j+j)$ and commute with the
energy and $\partial/\partial \phi$.  Thus we have $D=3N-3$ commuting
operators whose eigenvalues are constants of the motion in $D$
dimensions -- we may of course add ${\bf K}$ and get $3N$ constants of
the motion for our $N$ particles.  Thus our system is the quantum version of a system that is integrable by
Liouville's theorem.

The above integrals of motion can all be made up of the
hyper-angular-momenta components $q_a p_b-q_bp_a$.  Indeed the pattern
of these operators is as follows:  in 2 dimensions there is one
angular momentum operator which is conserved, i.e., commutes with the
Hamiltonian.  In 3 dimensions there are two extra angular operators
that are conserved but only the total angular momentum commutes with
the $2D$ angular momentum chosen to start with.  Likewise in 4
dimensions there are ${1 \over 2} 4 \times 3 = 6$ conserved angular
momentum operators -- 3 new ones -- and again it is the total angular
momentum that commutes with both the $3D$ total and the $2D$.  In $D$
dimensions there are ${1 \over 2} D(D-1)$ conserved angular momentum
operators of which ${1 \over 2} D(D-1) - {1 \over 2} (D-1)(D-2) = D-1$
are new ones that did not occur in the $D-1$ dimensional case.  Of
these the grand total sum of ${1 \over 2} D(D-1)$ squares is conserved
and commutes with all former totals.  Thus in $D$ dimensions there are
$D-1$ independent mutually commuting angular momenta and these,
together with the energy, give us the $D$ commuting operators expected
from Liouville's theorem.  We note that we could have chosen any pole
for our spherical polar coordinates; each choice gives us a different
complete set of operators.  

Of course the Keplerian case $V = -kM^2 r^{-1}$ also separates in
hyper-parabolic and hyper-spheroidal coordinates.  For such systems we
have a conserved Hamilton eccentricity vector (often called the
Runge-Lenz vector in atomic physics) c.f. Lynden-Bell \& Lynden-Bell
(1999), equation (2.25K)
$$e_a = k^{-1} M^{-3} \sum_b (q_ap_b-q_bp_a)p_b-q_a/r \ . $$

\label{lastpage}
\end{document}